\documentclass[useAMS,usenatbib]{mn2e}
\usepackage{epsfig}

\usepackage[mediumspace,Gray,squaren]{SIunits} 

\title[Star formation in high-$\MakeLowercase{z}$ massive galaxies]
{Measuring star formation in high-$\MakeLowercase{z}$ massive galaxies: \\ A mid-infrared to submillimeter study of the GOODS NICMOS Survey sample}

\author[Viero, Moncelsi, et al.] {M.~P.~Viero$^{1,2}$\thanks{E-mail:
marco.viero@caltech.edu}, L.~Moncelsi$^{1,3}$, E.~Mentuch$^{4,2}$,
F.~Buitrago$^{5,6}$,  A.~E.~Bauer$^{7,5}$, \newauthor
 E.~L.~Chapin$^{8}$, C.~J.~Conselice$^{5}$, M.~J.~Devlin$^{9}$,
M.~Halpern$^{8}$, G.~Marsden$^{8}$, \newauthor
C.~B.~Netterfield$^{2,10}$,
 E.~Pascale$^{3}$, P.~G.
P\'erez-Gonz\'alez$^{11,12}$, M.~Rex$^{12}$, \newauthor D.~Scott$^{8}$, M.~W.~L.~Smith$^{3}$, M.~D.~P.~Truch$^{9}$, I.~Trujillo$^{13,14}$, D.~V.~Wiebe$^{8}$\\
$^{1}$ California Institute of Technology, 1200 E. California Blvd., Pasadena, CA\ 91125, USA \\
$^{2}$ Department of Astronomy \& Astrophysics, University of Toronto, 50 St. George Street, Toronto, ON\ M5S~3H4, Canada \\
$^{3}$ Department of Physics \& Astronomy, Cardiff University, 5 The Parade, Cardiff, CF24~3AA, UK \\
$^{4}$ Department of Physics \& Astronomy, McMaster University, Hamilton, ON L8S~4M1 \\
$^{5}$ School of Physics and Astronomy, University of Nottingham, NG1 3AL, UK \\
$^{6}$ SUPA\thanks{Scottish Universities Physics Alliance}, Institute for Astronomy, University of Edinburgh, Royal Observatory, Edinburgh, EH9~3HJ, UK \\
$^{7}$ Australian Astronomical Observatory, PO Box 296, Epping, NSW 1710, Australia \\
$^{8}$ Department of Physics \& Astronomy, University of British Columbia, 6224 Agricultural Road, Vancouver, BC V6T~1Z1, Canada \\
$^{9}$ Department of Physics \& Astronomy, University of Pennsylvania, 209 South 33rd Street, Philadelphia, PA 19104 \\
$^{10}$ Department of Physics, University of Toronto, 60 St.George Street, Toronto, ON M5S~1A7 \\
$^{11}$ Departamento de Astrof\'{\i}sica, Facultad de CC. F\'{\i}sicas, Universidad Complutense de Madrid, E-28040 Madrid, Spain \\
$^{12}$ Steward Observatory, The University of Arizona, 933 N. Cherry Ave., Tucson, AZ 85721, USA \\
$^{13}$ Instituto de Astrof\'{\i}sica de Canarias, E-38205, La Laguna, Tenerife, Spain \\
$^{14}$ Departamento de Astrof\'isica, Universidad de La Laguna,
E-38205 La Laguna, Tenerife, Spain}

\begin{document}

\date{Accepted . Received 2011 April 18; in original form }

\pagerange{\pageref{firstpage}--\pageref{lastpage}} \pubyear{2009}

\maketitle

\label{firstpage}

\begin{abstract}
We present measurements of the mean
mid-infrared--to--submillimeter flux densities of massive
($M_{\star} \ga 10^{11}$\,$\rm M_{\sun}$) galaxies at
redshifts $1.7 < z < 2.9$, obtained by stacking positions of known
objects taken from the GOODS NICMOS Survey (GNS) catalog on maps at:
 24\,$\micro$m (\emph{Spitzer}/MIPS); 70, 100, and 160\,$\micro$m
(\emph{Herschel}/PACS); 250, 350, and 500\,$\micro$m (BLAST); and
870\,$\micro$m (LABOCA). A modified blackbody spectrum fit to the
stacked flux densities indicates a median [interquartile]
star-formation rate of $\rm SFR=63~[48, 81]$\,$\rm
M_{\sun}$\,yr$^{-1}$.
We note that not properly accounting for correlations between
bands when fitting stacked data can significantly bias the result.
The galaxies are divided into two groups,
disk-like and spheroid-like, according to their S\'{e}rsic
indices, $n$. We find evidence that most of the star formation is
occurring in $n \leq 2$ (disk-like) galaxies, with median
[interquartile] $\rm SFR=122~[100, 150]$\,$\rm
M_{\sun}$\,yr$^{-1}$, while there are indications that the $n
> 2$ (spheroid-like) population may be forming stars at a median
[interquartile] $\rm SFR=14~[9, 20]$\,$\rm M_{\sun}$\,yr$^{-1}$, if at all.
Finally, we show that star formation is
a plausible mechanism for size evolution in this population
as a whole,
but find only marginal evidence that it is what drives the expansion
of the spheroid-like galaxies.
\end{abstract}


\section{Introduction}

The observed structural properties of massive galaxies ($M_{\star}
\ga 10^{11}$\,$\rm M_{\sun}$) at high redshift ($z \ga 1$) are
difficult to reconcile with those of galaxies that populate the
local Universe. Most strikingly, they are on average much more compact in
size than local galaxies of similar mass \citep{daddi2005,
trujillo2006}. For the spheroid-like galaxy population, this size
evolution has been particularly dramatic \citep[a factor of 4--5
since $z\sim2$, see e.g.,][]{trujillo2007, buitrago2008,
damjanov2009}, with subsequent observations confirming these
findings \citep[e.g.,][]{Muzzin2009,trujillo2011}.
Only a tiny fraction of massive galaxies in the local Universe have sizes
comparable to those found at high redshift \citep{trujillo2009}.
The absence of similar mass counterparts in the local Universe
\citep{trujillo2009} implies that some mechanism is acting on
those high-redshift galaxies to make them grow in size
\citep{hopkins2009,bezanson2009}.

In order to understand the mechanism responsible for this galaxy
growth, a crucial point that needs to be addressed is the level of
star formation (or star-formation rate: SFR) in this population.
From an observational point of view, evidence for star formation
in massive galaxies at high redshift is unclear, especially for
the spheroid-like population. For example, small samples of
high-quality spectroscopy \citep{kriek2006,kriek2009a} find little
or no star formation in this population; whereas, about $50 \%$ of
these galaxies appear to have 24\,$\micro$m counterparts
\citep{perez2008b}, indicating an elevated level of star
formation. This discrepancy may be due to biases inherent to their
respective SFR estimators, which are either susceptible to errors
in extinction correction and require deep spectroscopic
observations; or probe emission from polycyclic aromatic
hydrocarbons (PAHs), and thus provide a poor constraint on the
thermal spectral energy distribution (SED).

An alternative probe of star formation is to observe in the
far-infrared/submillimeter bands (FIR/submm), where emission is
primarily from heated dust. It is known that in the local Universe
the dust luminosity in star-forming regions is correlated with SFR
\citep[e.g.,][]{Kennicutt98,Chary2001,Buat2007}, with the most
actively star-forming galaxies often the most dust obscured or
even optically thick in the optical/UV \citep{genzel1998}.
Therefore, it is reasonable to expect that if high-redshift,
compact, massive galaxies are vigorously forming stars, then they
should be observable in the rest-frame FIR/submm.

However, due to the large beams of current submm telescopes,
source confusion and flux boosting present significant obstacles
to studying the star formation properties of anything other than
the most luminous galaxies at high redshift \citep{Moncelsi2011}.
For example, the 1-$\sigma$ noise floor due to confusion in the 250\,$\micro$m band of
\emph{Herschel}/SPIRE is 5.8\,mJy \citep{nguyen2010}, which
corresponds to the flux from galaxies at $z\sim 2$ with bolometric
FIR luminosities of $L_{\rm FIR}\sim 2\times 10^{12}$\,$\rm
L_{\sun}$, i.e., ultra-luminous infrared galaxies (ULIRGs). As a
result, a catalog of galaxies at $z > 2$ robustly detected above
the confusion noise (5-$\sigma$) in the submm can only probe the
bright end of the luminosity distribution. Stacking provides a
mechanism to examine the full distribution, provided a reliable
external catalog extending to faint fluxes is available
\citep{Marsden2009,Pascale2009}.

\begin{figure}
\centering \vspace{0.2cm} \epsfig{figure=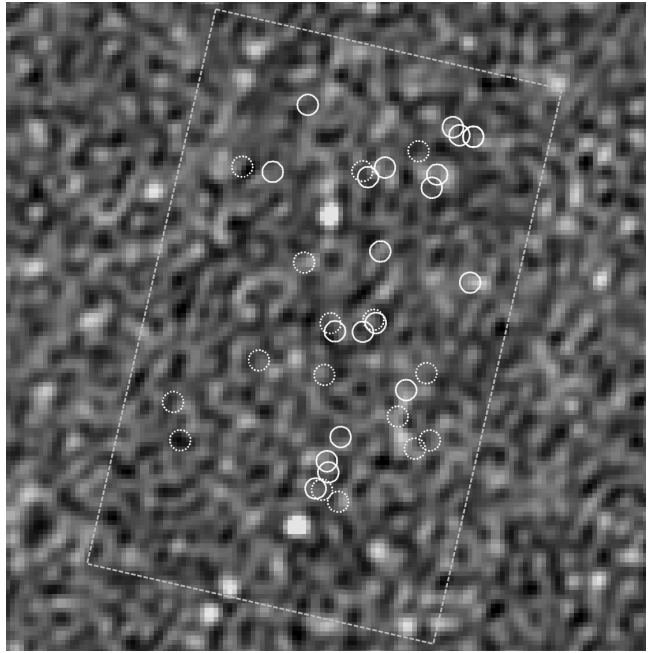, width=8.5cm}
\caption{GNS catalog positions (white circles, 36\arcsec\ in
diameter, solid are $n \le 2$; dotted are $n > 2$) overlaid on a
$20\arcmin \times 20\arcmin$ region of the BLAST 250\,$\micro$m map
in GOODS-South.  The overlapping \emph{Herschel}/PACS region is outlined as a dashed box.
The map has been convolved with a matched-filter
\citep[see][]{Chapin2011} to help enhance the regions of submm
emission. Most of the sources in our catalog lie along regions of
faint emission. Note that the BLAST beam is many ($\sim 18$--$30$)
times larger than a resolved galaxy, necessitating the stack.
Furthermore, since the angular resolution of \emph{Herschel}/SPIRE
images will only improve by a factor of two, stacking will still
be required to understand the FIR/submm properties of the faint
population.} \label{fig:blast_map}
\end{figure}

In this work we perform a stacking analysis using a catalog of
distant massive galaxies from the GOODS NICMOS Survey
\citep[GNS;][]{conselice2011} --- which we select to have stellar
masses $M_{\star}\ge 10^{11}$\,$\rm M_{\sun}$ and redshifts
$1.7<z<2.9$ --- on maps from: \emph{Spitzer}/MIPS
\citep{Rieke2004} at 24\,$\micro$m; \emph{Herschel}/PACS
\citep{poglitsch2010} at 70, 100, and 160\,$\micro$m; the
Balloon-borne Large Aperture Submillimeter Telescope
\citep[BLAST;][]{Devlin2009} at 250, 350, and 500\,$\micro$m; and the
Large APEX Bolometer Camera \citep[LABOCA;][]{Weiss2009} at
870\,$\micro$m. Our objective is to estimate the average SFRs of
high-redshift massive galaxies, and to look for differences
between the disk-like and spheroid-like galaxies. An alternative
approach, based on counterpart identification of similar GNS
catalog sources, is carried out by \citet{Cava2010}.

When required, we adopt the concordance model, a flat $\Lambda$CDM
cosmology with $\Omega_{\rm M}=0.274$, $\Omega_{\Lambda}=0.726$,
$H_0 = 70.5$\,km s$^{-1}$\,Mpc$^{-1}$, and $\sigma_8 = 0.81$
\citep{Hinshaw2009}.

\section{Data}

We perform our analysis on the Great Observatories Origins Deep
Survey South field (GOODS-South), also known as the Extended
Chandra Deep Field South (E-CDFS), which has field center
coordinates $\rm 3^{h}32^m30^s, -27^{\circ}48\arcmin 20\arcsec$.
Here we briefly describe the catalog and maps.

\subsection{Mass-Selected Catalog}\label{sec:mass_selected_catalog}

Our catalog is the \citet{buitrago2008} subset of the publicly
available GOODS NICMOS
Survey\footnote{\tt{http://www.nottingham.ac.uk/astronomy/gns/index.html}}
\citep{conselice2011}. Here we summarize its main features; for a
more detailed description see \citet{buitrago2008},
\citet{bluck2009} and \citet{conselice2011}.
For details concerning the data reduction procedure see \citet{magee2007}.
The GNS is a large \emph{HST} NICMOS-3 camera program of 60 $H$-band pointings
(180 orbits), with limiting magnitudes of $H\sim 26.8$ (5-$\sigma$),
optimized to collect data for as many massive ($M_{\star} \ga
10^{11}$\,$\rm M_{\sun}$) galaxies as possible at high redshift
($1.7 < z < 2.9$), making it the largest sample of such galaxies
to date. Of these, 36 are in the southern field for which we have
infrared and submm maps.

Redshifts and stellar masses of these objects are calculated
using the BVRIizJHK filters. Photometric redshifts are found using
standard techniques \citep[e.g.,][]{conselice2007}, while
spectroscopic redshifts for 7 objects are compiled from the
literature \citep{wuyts2008, popesso2009,balestra2010}. Stellar masses of these objects are estimated by
fitting the multi-color photometry to model SEDs --- produced with
stellar population synthesis models --- resulting in uncertainties
of $\sim 0.2$\,dex \citep[e.g.,][]{bundy2006}.

Additionally, due to the excellent depth and resolution of the
NICMOS images (pixel scale after resampling of $0.\arcsec
1$\,pixel$^{-1}$, and a point spread function [PSF] of $0.\arcsec
3$ full width half maximum [FWHM]), we are able to estimate the
S\'{e}rsic \citeyearpar{sersic1968} indices and sizes of the
objects using the {\tt GALFIT} code \citep{peng2002}. Average
properties of the sources used in our analysis are listed in
Table~\ref{tab:source_properties}.

\begin{table*}
 \centering
 \begin{tabular}{c|c|c|c|c|c|c|c|c|c}
  \hline
 $ $  &  N  & $z_{\rm median}$ & $z_{\rm iqr}$ & $M_{\star}$         & $R_{\rm e}$ & $n $ & $T$ & $L_{\rm FIR}$ &  SFR \\
 $ $  & $ $ & $ $                           & $ $                  & ($\rm M_{\sun}$)& (kpc)             & $  $   & (K) & ($\rm L_{\sun}$) & ($\rm M_{\sun}$\,yr$^{-1}$) \\
 \hline
 All & 36  & 2.285 & 1.980--2.500 & $1.85\times 10^{11}$ & 2.00 & 2.03 & $29.4^{+1.4}_{-0.8}~[27.3, 31.6]$ & $6.2^{+1.1}_{-1.0}~[4.7, 8.0]\times 10^{11}$ & $63^{+11}_{-11}$~[48, 81]\\
 $n \le 2$ & 20  & 2.285 & 2.085--2.500 & $1.93\times 10^{11}$ & 2.43 & $ 1.05 $ & $32.6^{+1.0}_{-0.4}~[30.8, 34.6]$ & $12.0^{+1.4}_{-1.5}~[9.8, 14.8]\times 10^{11}$ & $122^{+15}_{-15}$~[100, 150]\\
 $n > 2$ & 16  & 2.270 & 1.865--2.625 & $1.74\times 10^{11}$ & 1.49 & $ 3.25 $ & $27.6^{+0.3}_{-7.6}~[24.2, 30.8]$ & $1.4^{+0.2}_{-0.8}~[0.9,2.0]\times 10^{11}$ & $14^{+2}_{-8}$~[9, 20]\\
 \hline
 \end{tabular}
 \caption{Average properties of stacked samples. $R_{\rm e}$ is the
effective radius. Dust temperatures, bolometric FIR luminosities
and SFRs, corrected to a Chabrier \citeyearpar{chabrier2003} IMF,
are shown with the corresponding upper and lower Gaussian
uncertainties, and interquartile ranges in square brackets (see
Section~\ref{sec:results} for details).}
 \label{tab:source_properties}
\end{table*}

The selection for the GNS galaxies is based on mass and redshift,
with $1.7 < z < 2.8$.  These galaxies were located initially through color selection
techniques, such as the BzK  \citep{daddi2007}, ERO  \citep{yan2004} and
DRG \citet{papovich2006} criteria, and later refined through spectroscopic and
photometric redshifts within the two GOODS fields.  \citet{conselice2011}
perform several tests to ensure that the sample is complete.
A possible bias might be that extremely dusty galaxies could be missed by this criteria
due to attenuation, but the deep limiting $H$-band magnitude greatly exceeds
that of the expected upper bound for dusty SMGs \citep[$\sim
23.3$\,mag,][]{frayer2004}, so that we are confident that we are
not missing the dustiest galaxies due to attenuation.
 Lastly, it is expected that this selection of galaxies closely approximates
the true ratio of red to blue galaxies in these mass and redshift
ranges.
For more details concerning the selection technique and possible
biases \citep[see][]{conselice2011}.

\subsection{Spitzer}

We use the publicly available \emph{Spitzer}/MIPS map at
24\,$\micro$m from the Far Infrared Deep Extragalactic Legacy Survey
 (FIDEL \footnote{\tt{http://data.spitzer.caltech.edu/popular/fidel/20070917\_enhanced/docs/fidel\_dr2.html/}}).
The 5-$\sigma$ point source sensitivity of this map is 0.03\,mJy.

\subsection{PACS}

We use publicly available \emph{Herschel}/PACS
\citep{poglitsch2010} observations of the GOODS-South field from
the PACS Evolutionary Probe\footnote{\tt{http://www.mpe.mpg.de/ir/Research/PEP/}}
\citep[PEP;][]{Lutz2011} survey. The data was re-processed with the
\emph{Herschel} Processing Environment \citep[HIPE;][continuous integration build number 6.0.2110]{Ott2010}.
PEP was designed to provide data in all three PACS bands. Since PACS
can only observe in two bands simultaneously --- at 160\,$\micro$m
(red)  and either 70 (blue) or 100\,$\micro$m (green) --- we use two
sets of observations to produce maps at all three wavelengths. We
combine the available deep observations using the standard PACS
pipeline, choosing a high-pass filter parameter of 20 for the blue
and green bands, and 30 for the red band \citep[corresponding to
suppression of scales larger than 40 and 60\arcsec\ on the sky,
respectively; see][]{Muller2011}. In order to prevent ringing
effects around bright sources caused by the high-pass filter the
pipeline performs an initial crude reduction and automatically
masks out the brightest sources in the subsequent iterations of
de-glitching and filtering. The r.m.s.\@ depths of the final maps
are 0.31, 0.44, and 1.5\,mJy at 70, 100, and 160\,$\micro$m,
respectively.

As reported by \citet{Muller2011}, the relatively strong high-pass
filter adopted along with the masking of the bright sources may
attenuate the final photometry of faint sources. To account for
these effects, we produce maps of a few, isolated, unmasked, faint
point sources of different flux density, using the same parameters
as were used in the reduction of the GOODS-South maps; we then
mask these sources and create new maps. We use the average ratio
of the flux densities of the same sources in the two maps as our
estimate of the attenuation factor due to the high-pass filter. We
find that the magnitude of the attenuation mildly increases for
increasing wavelengths, as expected given the shape of the 1/f
noise over the relevant frequency range \citep[$\propto
f^{-0.5}$;][]{Lutz2011}. The estimated attenuation factors are
0.80, 0.78, and 0.75 at 70, 100, and 160\,$\micro$m, respectively.
Note that a slightly different approach was followed by
\citet{Lutz2011}, who perform tests on the red band by adding
simulated sources to the timelines before masking and high-pass
filtering; they find that the filtering modifies the fluxes by
16\% for very faint unmasked point sources. Despite the slight
disagreement with our finding at 160\,$\micro$m, and because of the
lack of an estimate for the blue and green bands from the PEP
team, we choose to adopt our three estimated factors for
consistency.

\subsection{BLAST}

The BLAST maps in GOODS-South\footnote{Available at:
\tt{http://blastexperiment.info/results.php}} consist of a deep
region covering $\sim 0.9$\,deg$^2$ which completely encompasses
the southern sources in the \citet{buitrago2008} catalog
(Figure~\ref{fig:blast_map}), and have r.m.s.\@ depths of 11, 9,
and 6\,mJy, at 250, 350, and 500\,$\micro$m, respectively
\citep{Devlin2009}. Due to large instrumental beams (36, 42, and
60\arcsec) and steep source counts \citep[approximately following
$dN/dS \propto S^{-3}$;][]{Patanchon2009}, source confusion
contributes substantially to the noise in these maps, and is
estimated to be $\sigma_{\rm confusion} \approx$ 21, 17, and
15\,mJy in the three bands \citep{Marsden2009}. The BLAST maps
were made using a naive mapmaker \citep{Pascale2011}. Further
details on the instrument may be found in \citet{Pascale2008},
while flight performance and calibration are provided in
\citet{Truch2009}.

\subsection{LABOCA}

The LABOCA E-CDFS Submm Survey \citep[LESS;][]{Weiss2009} provides
deep 870\,$\micro$m data, with an r.m.s.\@ depth to better than
1.2\,mJy across the full $30\arcmin \times 30\arcmin$ field, with
an effective resolution of 27\arcsec\ FWHM. For a detailed
description of the instrument see \citet{siringo2009}.

\section{Method}

\subsection{Stacking Formalism}\label{sec:method}

Stacking is a well established technique for finding the average
properties of objects which individually are undetectable by using
external knowledge of their positions in a map
\citep[e.g.,][]{Dole2006, wang2006, Marsden2009, Pascale2009}
. We follow the formalism of \citet[][hereafter
M09]{Marsden2009}, to which we refer to for a full description of
the stacking method. Here we summarize the salient features of the
technique.

\begin{figure}
\centering \vspace{0.2cm}\epsfig{figure=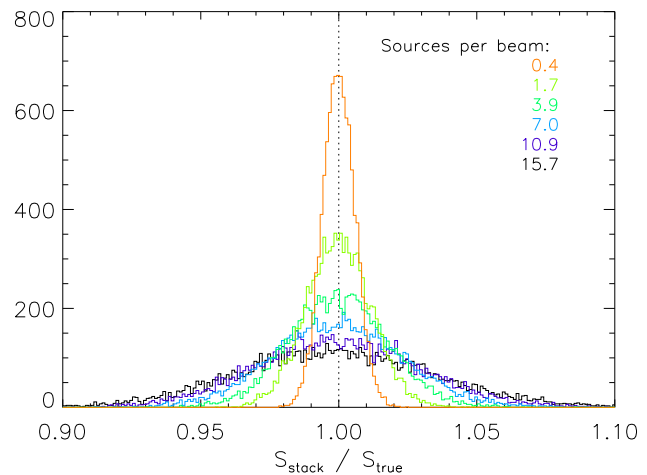, width=8.5cm}
\caption{Histograms showing the ratio of recovered stacked fluxes
to true flux for 10{,}000 simulations. The stacks were performed
on simulated 0.25\,deg$^2$ maps based on a random catalog of
12{,}500 sources, with size and source densities typical for deep
24\,$\micro$m MIPS catalogs. We have repeated the test for six beam
sizes in the range 10--60\arcsec, which probe the effects of
stacking at source densities ranging from 0.4 to 16 sources per
beam. As described in Section~\ref{sec:method} and in M09, larger
beams lead to larger uncertainties, but in all cases, the stacked
values are consistent with the true catalog flux, showing that
there is no bias when stacking on uncorrelated catalogs.}
\label{fig:simstack}
\end{figure}

M09 showed that the mean flux density of an external catalog is
simply the covariance of the mean-subtracted map with the catalog,
divided by the variance of the catalog density. If the catalog is
Poisson-distributed, then a powerful diagnostic is that the variance of
the source density should equal the mean, and the average flux density can be
re-written as the mean map value at the position of each catalog
source. This is true no matter what the size of the beam or
surface density of sources in the map, so long as the sources are
uncorrelated at the scale of the beam. The algorithm has been
extensively tested with Monte Carlo simulations on mock random
maps with increasing source densities, and was shown consistently
to recover the correct mean flux density, with no dependence on
the number of sources per beam (Figure~\ref{fig:simstack}). If
however the catalog \textit{is} clustered on the beam scale, the
stacked flux will be biased high, compared to the properly
normalized covariance, by a factor equal to the catalog variance
at the beam scale divided by the mean source density. In the
following section we show that this factor is consistent with
unity for our data.

Uncertainties and possible biases of our measurement are estimated
by generating random catalogs and stacking them on the actual maps
themselves. We find that the uncertainties are
Gaussian-distributed and scale as the map r.m.s.\@ (including
confusion noise) divided by the square root of the number of
catalog entries. Note that these uncertainties account for both
instrumental and source confusion noise, as well as for any
pixel-pixel correlations introduced by the map-making algorithm
(e.g., the \lq\lq drizzling\rq\rq\ technique used to produce the PACS maps
with the standard pipeline).

\subsection{Testing the Poisson Hypothesis}

Stacking provides an unbiased estimate of the mean flux only when
the sources in the sky are uncorrelated. While massive galaxies
have been shown to cluster quite strongly
\citep[e.g.,][]{foucaud2010}, we find that on scales relevant for
this analysis they are essentially Poisson-distributed, as we show
with the following tests:

1) In the presence of clustering, the FWHM of the postage-stamp of
stacked sources would be larger than the nominal instrumental PSF.
We compare our measured stacked 24\,$\micro$m PSF to that measured
from stacking the sources used in M09 \citep{Magnelli2009}, which
were shown to be Poisson distributed (see Figure~3 of M09), and find
that they are identical.

2) If the sources are Poisson-distributed over a given scale, then
by definition the average number of sources in a cell of that size
should equal the variance. We test that by dividing the field into
equal sized cells, from 2.7 to 0.225\arcmin\ on a side, and find
that the ratio of the variance to the mean is consistent with
unity at all scales.

3) In the presence of strong clustering around massive galaxies we
would expect to find more sources per beam surrounding the
galaxies than would be found at random. We calculate the number of
sources inside a BLAST beam radius at the locations of each
massive galaxy and compare that to what we would expect at random.
From 1{,}000 Monte Carlo simulations we find $1.10\pm 0.13$,
$1.16\pm 0.17$, and $1.28\pm 0.21$ sources per beam at 250, 350,
and 500\,$\micro$m, compared to the measured 1.04, 1.13, and 1.17,
respectively. We extend this test to galaxies with
log$(M_{\star}/ \rm M_{\sun})> 9$, to account for the possibility of
less massive galaxies clustering around our more massive ones. We
find there are $2.85\pm 0.40$, $3.83\pm 0.51$, and $5.97\pm 0.73$
sources per beam at 250, 350, and 500\,$\micro$m, compared to the
measured 2.53, 4.04, and 5.87, respectively. Thus, while there are
multiple sources per beam at all wavelengths, because their
distribution is consistent with Poisson, they do not bias the
result.

There still remains the possibility, however, that even fainter
($< 13$\,$\micro$Jy at 24\,$\micro$m), undetected sources cluster around
detected ones. We can estimate their potential contribution in the
following way. If clustered, faint sources contribute
significantly to the stacked flux density for large beams then after
convolving the 24\,$\micro$m map (whose beam is 6\arcsec) with a much
larger beam we would expect the stacked flux density to increase.
On the other hand, as described in the previous section, if the
faint sources are Poisson-distributed then we would expect only
the noise to increase. We find that after convolving the
24\,$\micro$m map with a 60\arcsec\ beam, the stacked flux density
per source is $0.08\pm 0.11$\,mJy, compared to the original
$0.081\pm 0.005$\,mJy (see Table~\ref{tab:stack_results}). Thus,
the stacked signal does not change, but the errors increase
substantially, which is consistent with what we would expect from
additional, Poisson-distributed sources in the beam. We therefore
conclude that the contribution from faint clustered sources is
negligible.

\subsection{SED Fitting, IR Luminosities, and Star-Formation Rates}\label{sec:fitting}

We model the thermal dust emission as a modified black body with
an SED of the form;
\begin{equation}
S_{\nu} = A \nu^{\beta} B(\nu,T),
\end{equation}
where $B(\nu,T)$ is the blackbody spectrum with amplitude $A$,
and $\beta$ is the emissivity index, which effectively takes into account the
variability of dust temperatures within a single galaxy.  Following \citet{Blain2003}
we set $\beta$ to 1.5.  Additionally, we replace the mid-infrared
exponential on the Wien side of the spectrum with a power-law of
the form $f_{\nu} \propto \nu^{-\alpha}$ \citep[with $\alpha = 2$,
following][]{Blain2003}, which in practice means imposing that the two functions and their first
derivatives be equal at the transition frequency.  In our case that transition occurs at rest-frame $\sim 74\, \micro$m 
(for $T \sim 30\,$K; see \S~\ref{sec:bestfits}).

Our SED fitting procedure estimates the amplitude and temperature of the
above template, keeping $\alpha$ and $\beta$ fixed. For the BLAST points, the SED fitting procedure \citep[described
in detail in][]{Chapin2008} takes the width and shape of the
photometric bands into account, as well as the absolute
photometric calibration uncertainty in each band
\citep[see][]{Truch2009}. Correlations due to instrumental noise
are estimated and accounted for with a Monte Carlo procedure.
Because we do not possess similar detailed data for {\sl
Spitzer}/MIPS and LABOCA, these photometric points are not
color-corrected, whereas we do apply a color-correction to the
PACS points, following the standard procedure described in
\citet[][see their Table 4.2, for a power law
$\nu^{-2}$]{Muller2011}; the color-correction factors are 1.016,
1.012, 1.017 at 70, 100, and 160 um, respectively, and have a
negligible impact on the final results. The PACS points are
assumed to have completely uncorrelated instrumental noise among
bands.

Correlated confusion noise must also be accounted for in the fit,
as these correlations reduce the significance of a combination of
single band detections. We estimate the Pearson coefficients of
the correlation matrix for all bands (see
Table~\ref{tab:covariance_matrix}) from the beam-convolved maps 
within a region of 0.064\,deg$^2$ that encompasses all the sources
in the GOODS-South NICMOS catalog. In \S~\ref{sec:bestfits} we will show how the effect of correlations
between bands is quite significant, especially among PACS and BLAST
bands \citep[see also][]{Moncelsi2011}, and thus including them in
the SED fitting algorithm is crucial.  

\begin{table}
  \centering
  \begin{tabular}{ccccccccc}
   \hline
   Band &   &      &      &      &       &      &        & \\
   ($\micro$m) & 24 & 70 & 100 & 160 & 250 & 350 & 500 & 870\\
   \hline
  24  & 1 & 0.11 & 0.13 & 0.23 &  0.35 & 0.28 &  0.22  &  0.05 \\
70  &   & 1    & 0.92 & 0.77 &  0.22 & 0.15 &  0.08 & 0.006 \\
100 &   &      & 1    & 0.86 &  0.27 & 0.19 &  0.11 & 0.007  \\
160 &   &      &      & 1    &  0.44 & 0.33 & 0.20 & 0.04  \\
250 &   &      &      &      &     1 & 0.70 &  0.62  &  0.11 \\
350 &   &      &      &      &       & 1    &  0.70  & 0.14  \\
500 &   &      &      &      &       &      &  1     & 0.13  \\
870 &   &      &      &      &       &      &        & 1      \\
   \hline
  \end{tabular}
  \caption{Pearson correlation matrix for all bands.}
  \label{tab:covariance_matrix}
\end{table}


SEDs are corrected for redshift by assuming the median redshift
for each subset (see column 3, Table~\ref{tab:source_properties}).
Interquartile errors reflecting the uncertainty in dimming due to
the width of the redshift bin are estimated with a Monte Carlo,
where 1000 mock redshifts with the same distribution as the chosen
subset (i.e., all, disk-like, and spheroid-like) are drawn, and
the dimming factor for each redshift is calculated.

The resulting infrared luminosity, $L_{\rm FIR}$, is
conventionally the integral of the rest-frame SED between 8 and
1000\,$\micro$m, and the SFR is estimated as
\begin{equation}
{\rm SFR}~[{\rm M}_{\sun}\,{\rm yr}^{-1}] = 1.728 \times 10^{-10}
\times L_{\rm FIR}~ [\rm L_{\sun}],
\end{equation}
from \citet{Kennicutt98}, which assumes a Salpeter
\citeyearpar{Salpeter1955} initial mass function (IMF). To convert
to a Chabrier \citeyearpar{chabrier2003} IMF, log(SFR) must be
corrected by lowering 0.23\,dex \citep[e.g.,][]{kriek2009b,
vandokkum2010}.

\section{Results}\label{sec:results}

\subsection{Stacking Results}\label{sec:stacking_results}

Stacked flux densities and 1-$\sigma$ uncertainties are reported in the
second column of Table~\ref{tab:stack_results}. We find
statistically significant, non-zero signals in all the submm
bands, with 2-, 3-, 3-, and 4-$\sigma$ detections at 250, 350,
500, and 870\,$\micro$m, respectively, as well as robust 16-, 3-, 4-,
4-$\sigma$ detections at 24, 70, 100, 160\,$\micro$m, respectively.

\begin{table}
  \centering
  \begin{tabular}{ccccc}
   \hline
   Band & All & $n \le 2$ (disk-like) & $n > 2$ (spheroid-like) \\
     $(\micro$m) &  (mJy/source)&  (mJy/source)&  (mJy/source)\\
   \hline
     24 &  $0.081 \pm  0.005 $ & $0.130 \pm  0.007 $ & $ 0.020 \pm 0.007 $ \\
     70 &  $0.16 \pm  0.07 $ & $0.36 \pm  0.09 $ & $ -0.05 \pm 0.10 $ \\
     100 & $ 0.39 \pm 0.09 $ & $ 0.84 \pm 0.13 $ & $-0.17 \pm 0.14  $\\
     160 &  $1.2 \pm  0.3 $ & $2.9 \pm  0.5 $ & $ -0.66 \pm 0.50 $ \\
     250 &  $5.0 \pm  2.9 $ & $9.3 \pm  3.9 $ & $ -0.3 \pm 4.4 $ \\
     350 &  $7.9 \pm  2.3 $ & $10.7 \pm  3.1 $ & $  4.5 \pm  3.5 $ \\
     500 &  $5.3 \pm  1.9 $ & $6.2 \pm  2.6$ & $  4.2 \pm  2.9 $ \\
     870 &  $0.97 \pm  0.26 $ & $1.03 \pm  0.35$ & $   0.9 \pm 0.4 $ \\
   \hline
  \end{tabular}
  \caption{The mean flux densities of massive galaxies in the GNS catalog from stacking. Reported are the results for all of the sources, as well as those identified as disk-like and spheroid-like, based on their S\'{e}rsic indices,
  $n$.}
  \label{tab:stack_results}
\end{table}


Next, we divide the catalog by S\'{e}rsic index into: those with
$n > 2$, which are spheroid-like and thus more likely to have
suppressed star formation; and those with $n \le 2$, which are
disk-like and thus more likely to be actively forming stars
\citep{ravindranath2004}.
Contamination of one population into the other has been shown with simulations to be very low  \citep[$< 10\% $; ][]{buitrago2011}, but when galaxies do cross the $n  = 2$ threshold, it is always from $n < 2$ to $n > 2$, i.e., from spheroid-like to disk-like.

The results are listed in the third and fourth columns of Table~\ref{tab:stack_results}.
At 24\,$\micro$m, we measure a distinct signal from both populations, with 19-$\sigma$ and
3-$\sigma$ detections from the disk-like and spheroid-like sources, respectively.
At longer wavelengths, for the disk-like
population we detect signals with greater significance than 
that of the combined catalog, between 2.5-$\sigma$ and 6.5-$\sigma$, in
each FIR/submm band; whereas for the spheroid-like population we
find a much weaker signal, with four bands consistent with zero.

While the error on the stacks is Gaussian, the uncertainty
associated with the average rest-frame $L_{\rm FIR}$ is dominated
by the width of the redshift distribution, which is not Gaussian.
Thus, for estimating $T$, $L_{\rm FIR}$, and SFR
(Section~\ref{sec:SEDs_and_SFRs}), we choose to adopt
the median value and interquartile range, as they best reflect the
asymmetric shape of the redshift distribution, which ultimately
determines the uncertainty of our measurement. For reference we also
quote the Gaussian uncertainties. We anticipate that the lower
Gaussian errors on $T$, $L_{\rm FIR}$, and SFR for the
spheroid-like subset exceed the lower bound of the interquartile
range, and reflect the elevated level of uncertainty in our
measurement.

\begin{figure*}
\centering \epsfig{figure=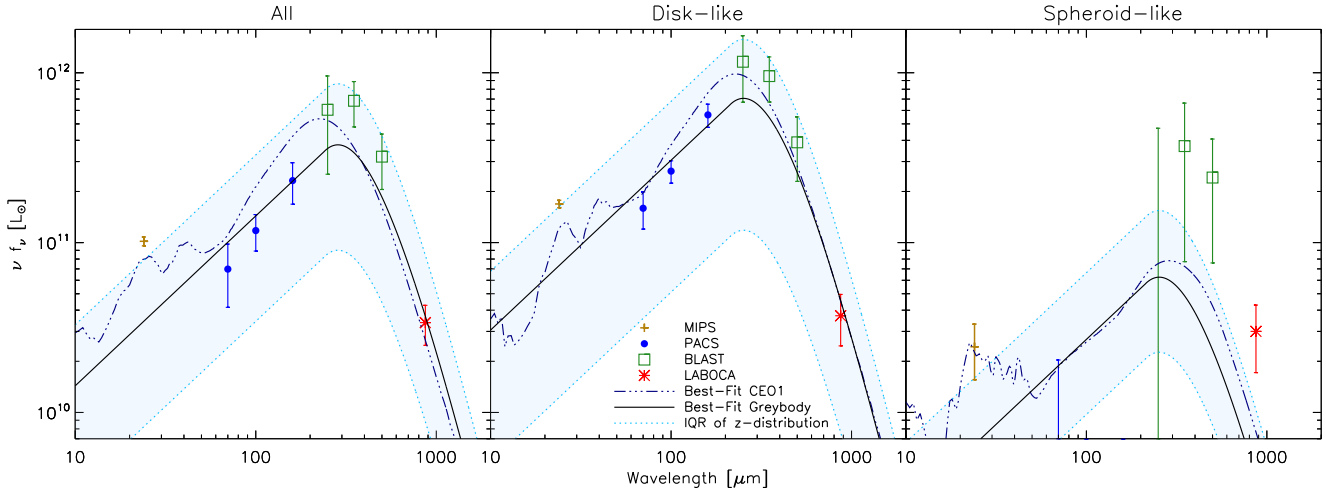, width=18cm} \caption{SED
fits to the stacked flux densities of all (left), disk-like
(center), and spheroid-like (right) sources. The median value of
the redshift distribution, $z\sim 2.3$, is used here to convert
flux densities into rest-frame luminosity. The brown crosses are
from Spitzer (24\,$\micro$m); the blue dots are from PACS (70, 100,
and 160\,$\micro$m); the green squares are from BLAST (250, 350, and
500\,$\micro$m); and the red asterisks are from LABOCA (870\,$\micro$m).
The error bars represent the 1-$\sigma$ Gaussian uncertainties
from the stacks as listed in Table~\ref{tab:stack_results}.  The
SED is modeled as a modified black body with a fixed emissivity
index $\beta = 1.5$, and a power-law approximation on the Wien
side with slope $\alpha=2$. The solid black lines are the best-fit
SEDs, while the dotted light-blue lines enclosing the shaded
regions show the uncertainties due to the width of the redshift
distribution (interquartile range), which clearly dominate over
the Gaussian errors on the stacks (see
Section~\ref{sec:stacking_results}). The navy 3-dot-dashed lines
are the best-fit, redshift-averaged templates from
\citet{Chary2001}.} \label{fig:sed}
\end{figure*}

\subsection{Contribution of stellar emission}
\label{sec:stellar}

At $z\sim2.3$ the observed 24\,$\micro$m band probes rest-frame
wavelengths of 6--7\,$\micro$m, which in addition to PAH emission, is
where the Rayleigh-Jeans tail of stellar emission lies. It is then 
plausible that the emission we find in this band may be entirely
attributed to stellar emission. Since our
detection of PAH and dust emission, particularly for the spheroid
population, is supported
strongly by this data point (given its high signal-to-noise), in this section,
we investigate the contribution of pure stellar emission to the observed
24\,$\micro$m band. 

Any additional emission not attributed to stellar emission
is likely associated with PAH emission, which in turn is accompanied
by longer wavelength dust emission that we have inferred our
SFRs from. We note that this emission may be
associated with either star forming regions or evolving main sequence
stars (such as AGB and TP-AGB stars). Typically, emission associated with star formation dominates in most
 galaxies (even those with moderate SFRs) as the infrared
light-to-mass ratio is up to three orders of magnitude larger for a simple stellar population (SSP) of $10^7\, \rm yrs$ compared
to an SSP of $10^9\, \rm yrs$ \citep{piovan2006}, where TP-AGB emission would be most significant.

To investigate the contribution of pure stellar emission in our
sample, we calculate the
predicted 24\,$\micro$m observed flux densities from stellar
population synthesis
models using redshifts and stellar masses as per our catalog (see
Section~\ref{sec:mass_selected_catalog}). We opt to use a galaxy
template with solar metallicity and an exponentially declining SFR
with an e-folding time of 500\,Myr, generated with the stellar
population synthesis code PEGASE.2 \citep{Fioc1997}. Output from
non-stellar emission or evolving main-sequence stars is not
included, as the source of non-stellar emission at 7\,$\micro$m is
assumed to be the same
as that of the FIR emission. Assuming a formation redshift of $z=9$,
the galaxy ages
range from 1.5 to 3\,Gyr and the predicted 24\,$\micro$m flux
densities due to stellar emission range from 1.3 to 8.8\,$\micro$Jy,
depending primarily on the galaxy's redshift. For each stacked
sample, we find the predicted contamination per galaxy from
stellar emission is 3.0, 2.9, and 3.9\,$\micro$Jy for the entire
sample, the disk-like and
spheroid-like populations, respectively. Contributing at most 20\%
(see Table~\ref{tab:stack_results}) to the observed 24\,$\micro$m flux
densities of the spheroid
population and less than 5\% to the observed 24\,$\micro$m flux
densities of the disk-like and total sample, we conclude that the 
mid-infrared observations (restframe 7-8\,$\micro$m) included in our
analysis are dominated by non-stellar emission (i.e., dust and PAH emission).
\subsection{Best-fit SEDs and Star-Formation Rates}\label{sec:SEDs_and_SFRs}
\label{sec:bestfits}
The best-fit SED and interquartile range to the stacked values of
the complete catalog are shown in the left panel of
Figure~\ref{fig:sed}, corresponding to a median (plus/minus
Gaussian) [interquartile] temperature of $T=
29.4^{+1.4}_{-0.8}~[27.3, 31.6]$\,K, luminosity of $L_{\rm
FIR}=6.2^{+1.1}_{-1.0}~[4.7, 8.0]\times 10^{11}$\,$\rm L_{\sun}$ , and
${\rm SFR}=63^{+11}_{-11}~[48, 81]$\,$\rm M_{\sun}$\,yr$^{-1}$.

As a sanity check, we compare our modified blackbody approximation 
to the best-fit template of \citet*[hereafter CE01]{Chary2001}.
The purpose of this is simply to reassure ourselves that an exponential
approximation on the Wien side of the thermal SED is not an unreasonable way to estimate the
contribution to the bolometric luminosity short of the SED peak, rather than
an attempt to derive SFRs from fitting SED templates.
Thus, for each of the 101 templates, we approximate the stacked SED by
taking the average of templates shifted to the redshift of each
galaxy in the catalog; this acts to smear out the otherwise
highly-variable PAH region of the rest-frame SED probed by the
24\,$\micro$m band. We fit the resulting template to our photometric
points without accounting for calibration uncertainties, color
corrections, or correlations among bands. The best-fit template is
shown as a 3-dot-dashed line in Figure~\ref{fig:sed}, and falls
well inside our error region. However, the SFR of the best-fit
template is ${\rm SFR}= 87$\,$\rm M_{\sun}$\,yr$^{-1}$, or
$\sim38$\% larger than our modified blackbody estimate. 
This overestimate likely arises because
the fit with the CE01 template does not include the substantial
correlations among bands (see Section~\ref{sec:fitting}), which
reduce the significance of the combination of individual
photometric points.

We then separately fit the stacked flux densities measured for
disk-like and spheroid-like galaxies.
The best-fit modified blackbody SED for the disk-like population
is shown in the center panel of Figure~\ref{fig:sed}, and results
in a median (plus/minus Gaussian) [interquartile] temperature of
$T = 32.6^{+1.0}_{-0.4}~[30.8, 34.6]$\,K, luminosity of $L_{\rm
FIR}=12.0^{+1.4}_{-1.5}~[9.8, 14.8]\times 10^{11} \rm L_{\sun}$,
and ${\rm SFR}= 122^{+15}_{-15}~[100, 150]$\,$\rm
M_{\sun}$\,yr$^{-1}$. The best-fit CE01 template is also shown,
and corresponds to a ${\rm SFR}= 142$\,$\rm M_{\sun}$\,yr$^{-1}$.

Likewise, the best-fit modified blackbody SED for the
spheroid-like population is shown in the right panel of
Figure~\ref{fig:sed}, and results in a median (plus/minus
Gaussian) [interquartile] temperature of $T =
27.6^{+0.3}_{-7.6}~[24.2, 30.8]$\,K, luminosity of $L_{\rm
FIR}=1.4^{+0.2}_{-0.8}~[0.9,2.0]\times 10^{11}~\rm L_{\sun}$, and
${\rm SFR}= 14^{+2}_{-8}~[9, 20]$\,$\rm M_{\sun}$\,yr$^{-1}$. Note
that the lower Gaussian errors exceed the lower bound of the
interquartile range, thus reflecting the elevated level of
uncertainty in our measurement. Once again, the best-fit CE01
template is shown, which corresponds to a ${\rm SFR}= 16$\,$\rm
M_{\sun}$\,yr$^{-1}$.

Finally, to check that contributions to the rest-frame SED from PAHs, which are highly variable,
are not significantly influencing the best-fit result, we re-fit the modified black body after excluding: i) just the 24\,$\mu$m point; and ii) all points below rest-frame 100\,$\mu$m.
In the first scenario we find ${\rm SFR}= 57^{+9}_{-14}~[42,72],  109^{+11}_{-18}~[91,135]$, and $12^{+5}_{-7}~[8,17]$\,$\rm M_{\sun}$\,yr$^{-1}$, while in the second scenario we find ${\rm SFR}= 67^{+11}_{-16}~[50,83],  129^{+16}_{-23}~[107,160]$, and $30^{+10}_{-7}~[19,42]$\,$\rm M_{\sun}$\,yr$^{-1}$, for all, disk-like, and spheroid-like galaxies, respectively.
In the first case, the SFRs decrease only marginally, and within the error bars, suggesting that the 24\,$\mu$m point alone does not unreasonably influence the result.
In the second case, SFRs for all and disk-like galaxies are mildly affected, while the spheroid-like galaxies are artificially boosted by a factor of two simply because we have removed the two data points consistent with zero.

Thus, although the best-fit SED to the combined stack returns a
robust, 4-$\sigma$ detection, it is clear that signal is dominated
by the disk-like, $n \le 2$ galaxies, which are detected at
5-$\sigma$. The best-fit to the spheroid-like, $n > 2$ galaxies,
on the other hand, returns a marginal 2-$\sigma$ result, which
suggests, but does not formally detect, a low level of star
formation taking place in the spheroid-like population.

We note that if correlations between bands are not properly accounted for when finding the best-fit SED, the corresponding SFRs are 94, 163, and 32\,$\rm M_{\sun}$\,yr$^{-1}$ for all, disk-like, and sphere-like galaxies, respectively. This is significantly different, and the reason is that if correlations between bands are not considered, more weight is attributed to the BLAST measurements than is appropriate, pulling the best-fit up.
Intuitively this makes sense: since confusion noise arises from multiple sources in a beam, a larger beam has more sources in it and thus more variance, i.e., more confusion noise.  Of course, for bands of similar wavelengths those sources are more or less present in each map, resulting in confusion noise that is not independent.
Though this will be less of a problem for \emph{Herschel}/SPIRE, the beam size and thus improvement in confusion noise is only of order $\sim 2\times$, so that correctly accounting for correlated confusion noise will still be very important.



\section{Discussion}

\subsection{Consequences for Galaxy Growth}

There are indications that massive galaxies at 
high redshift are the cores of present-day massive ellipticals
\citep{hopkins2009, bezanson2009}, and that the growth of these
galaxies takes place mostly in the outskirts via star formation
and minor mergers \citep[][]{hopkins2009, vandokkum2010} --- a
process sometimes referred to as \lq\lq inside-out\rq\rq\ growth,
which has also been observed in hydrodynamical cosmological
simulations \citep{Naab2009,Johansson2009,Oser2010}. Furthermore,
\citet{vandokkum2010} find that a SFR of $55\pm$13\,$\rm
M_{\sun}$\,yr$^{-1}$ at $z\sim 2$ is necessary to account for the
mass growth they observe in massive galaxies selected by number
density, from $z = 2$ to the present day, and that for $z \ga 1.5$
the mechanism for growth is primarily star formation.
Note that nearly half of their $z\sim2$ subsample of massive galaxies has
$n < 2$ (see right panel of their Figure~7) --- a fraction similar to our own.
Our measurement of $63~[48, 81]$\,$\rm M_{\sun}$\,yr$^{-1}$ for the
entire sample 
agrees well with their finding
; however, we do not find convincing evidence that star formation is the mechanism
driving the expansion in spheroid-like galaxies.

\subsection{Potential contribution from other sources of dust heating}

Star formation may not be the only explanation for infrared emission in
our sample which consists of very massive, yet relatively young systems.
The age of the universe by $z=3$--1.8 is just $\sim1.5$--3\,Gyr,
providing a strict upper limit on the ages of the stellar
populations. If these galaxies formed the bulk of their stellar
mass, as their colors suggest, early on, then it is likely that
they contain a large population of stars undergoing
post-main-sequence phases in which carbonaceous dusty material is
being produced and heated by very luminous stars. While it is
generally accepted in the current versions of stellar population
synthesis models \citep{Maraston2005,Bruzual2010,Conroy2010} that
thermally-pulsating asymptotic giant branch (TP-AGB) stars can contribute
up to 70\% of the emission seen in the near-infrared bands at ages
of 1--2 Gyr, there has been little work calibrating the global contribution of this
population to a galaxy's infrared luminosity. By extension, given the masses and ages of our
galaxies, we cannot rule out the possibility that the infrared
emission we have detected in our analysis is partially due to dust
heated and created by post-main-sequence stars.

\subsection{Red and Dead?}\label{sec:red_dead}
Our best-fit SED to stacked data does not  correspond to a  formal detection of star
formation in the spheroid-like ($n > 2$) galaxies,
however, the high 24\,$\micro$m flux might indicate a non-zero star-formation rate.
Though we have stated that 24\,$\micro$m emission alone is insufficient for
accurately estimating the \emph{level} of star formation in a galaxy,
locally, 24\,$\micro$m emission is typically well correlated
with star-forming regions \citep{Calzetti2007,Kennicutt2009}.
Additionally, emission from evolved stars seems unable to account
for the level of 24\,$\micro$m emission observed (\S~\ref{sec:stellar}).
Therefore, it seems plausible that star formation may be occurring in these galaxies at some level.
Furthermore, if a low level of star formation does indeed exist, given the noise properties of our maps,
the only bands which would permit a significant detection are the
24 and 870\,$\micro$m bands --- those in which our measurements
have signal-to-noise greater than 2.5.

If star formation is occurring in the spheroid-like galaxies, even at a low level, and if they are fair analogs
of the apparently red-and-dead compact spheroids seen by e.g., \citet{kriek2009b},
then why is it that star formation is not significant in ultra-deep
spectroscopy?  One possibility is that the star formation is localized in very dust-obscured regions.
Note that although \citet{kriek2009b} detect a faint
H$\alpha$ line, concluding that SFRs are at most 2--4\,$\rm
M_{\sun}$\,yr$^{-1}$, that is after correcting for a very
moderate amount of extinction (A$_{v} = 0$--0.3\,mag). For this
galaxy to actually be forming around 14\,$\rm
M_{\sun}$\,yr$^{-1}$, $L_{{\rm H}\alpha}$ would need to have been
underestimated by a factor of $\sim 3.5$--7, which corresponds
to  1.4--2.1\,mag of extinction.
Considering that resolved observations of nearby galaxies showing extinction values
of A$_{\rm H\alpha} > 3$ are common in H\,{\sc ii} regions
\citep{prescott2007} and regions of high star formation
\citep{mentuch2010}, this amount of extinction is not unrealistic.


Lastly, note that our low levels of observed star formation are in 
disagreement with \citet{Cava2010}, who (after correcting by
0.23\,dex due to differences in the assumed IMF) find an average SFR of
30--$60$\,$\rm M_{\sun}$\,yr$^{-1}$ for the spheroid-like galaxies.
Their average SFRs are based on photometry of individual
galaxies at 24\,$\micro$m, and at 250, 350, and 500\,$\micro$m from
\emph{Herschel}/SPIRE with a mean
detection fraction for the spheroid-like population of $\sim0.4$
at 24\,$\micro$m and $\sim0.15$ at 250\,$\micro$m.
This selection makes it difficult to properly compare measurements.

\section{Summary}

Our goal was to search for evidence of star formation in
high-redshift massive galaxies, with the hope of leading to a
better understanding of the mechanisms responsible for their
growth. We found that on average the full catalog of sources are
forming stars with a median [interquartile] ${\rm SFR}= 63~[48,
81]$\,$\rm M_{\sun}$\,yr$^{-1}$, which can be decomposed into a
relatively strong signal for the disk-like galaxies, with a median
[interquartile] ${\rm SFR}= 122~[100, 150]$\,$\rm M_{\sun}$\,yr$^{-1}$, and a
marginal signal for the spheroid-like population, with a median
[interquartile] ${\rm SFR}= 14~[9, 20]$\,$\rm
M_{\sun}$\,yr$^{-1}$.

The level of star-formation detected for the full catalog
is in good agreement with other measurements of galaxy growth
\citep[e.g.,][]{vandokkum2010} which show that star formation can account for
most of the growth at these redshifts.
However, despite having detected stacked emission at 24 and 870\,$\micro$m,
we are unable to say convincingly that star formation is responsible for the dramatic
size evolution of the spheroid-like population.

Lastly, though a red sequence appears to already be in place by
$z\sim 2$ \citep{kriek2009a}, we found hints that perhaps
the red, compact, spheroid-like galaxies may not be completely dead.
Future stacking work with larger catalogs and better maps will go
a long way to further understanding this question.
Better data bracketing the peak with SPIRE \citep[250, 350, and
500\,$\micro$m;][]{Griffin2010} will make more robust estimates
of the SED possible, and will greatly increase our understanding of star formation in
high-redshift massive galaxies.


\section*{Acknowledgments}

The authors would like to extend a big thanks to the LESS team for
providing the 870\,$\micro$m map, and to David Frayer for the \emph{Spitzer} 70\,$\micro$m
map used in the first version of this paper.
We would also like to thank Ian Smail, Kimberly
Scott, and Ivana Damjanov for constructive comments.
Finally, we would like to thank Herv\'e Aussel for his helpful advice. 
We acknowledge the support of NASA through grant numbers NAG5-12785, NAG5-13301, and NNGO-6GI11G, the NSF Office of Polar Programs, the Canadian Space Agency, the Natural Sciences and Engineering Research Council (NSERC) of Canada, and the UK Science and Technology Facilities Council (STFC). 
This publication is based on data acquired with the Atacama Pathfinder
Experiment (APEX) under programme numbers 078.F-9028(A),
079.F-9500(A), 080.A-3023(A), and 081.F-9500(A). APEX is a
collaboration between the Max-Planck-Institut fur Radioastronomie,
the European Southern Observatory, and the Onsala Space
Observatory.
\bibliographystyle{mn2e}
\bibliography{refs}

\label{lastpage}
\end{document}